\documentclass[print,structabstract]{aa}
\usepackage{amsmath}
\usepackage{graphicx}
\usepackage{txfonts}
\usepackage{natbib}
\bibpunct{(}{)}{;}{a}{}{,}
\usepackage{url}           

\usepackage{color}
\usepackage{ulem}
\definecolor{lightred}{rgb}{0.8,0.0,0.0}
\definecolor{lighterred}{rgb}{0.8,0.6,0.6}
\definecolor{lightgreen}{rgb}{0.1,0.6,0.1}

\begin{document}

\title{Revealing the nature of magnetic shadows with numerical 3D-MHD
  simulations.}
\titlerunning{The nature of magnetic shadows}

\author{C. Nutto \and O. Steiner \and M. Roth}

\institute{Kiepenheuer-Institut f\"ur Sonnenphysik, 
           Sch\"oneckstrasse 6,
           79104 Freiburg, Germany\\
           \email{nutto@kis.uni-freiburg.de}
           }

\abstract{}{We investigate the interaction of magneto-acoustic waves with 
magnetic network elements with the aim of finding possible signatures of 
the magnetic shadow phenomenon in the vicinity of network elements.}
{We carried out three-dimensional numerical simulations of magneto-acoustic wave 
propagation in a model solar atmosphere that is threaded by a 
complexly structured magnetic field, resembling that of a typical 
magnetic network element and of internetwork regions.
High-frequency waves of 10 mHz are excited at the bottom of the simulation 
domain. On their way through the upper convection zone and through the 
photosphere and the chromosphere they become perturbed, refracted, and 
converted into different mode types. We applied a standard Fourier analysis 
to produce oscillatory power-maps of the line-of-sight velocity.}
{In the power maps of the upper photosphere and the lower chromosphere, we clearly
see the magnetic shadow: a seam of suppressed power surrounding the magnetic
network elements. We demonstrate that this shadow is linked to the 
mode conversion process and that power maps at these height levels show the 
signature of three different magneto-acoustic wave modes.}{}


      \maketitle

\section{Introduction}\label{sec:introduction}

The feature-rich solar chromosphere
is one of the most disputed topics in present solar research \citep{rutten12}. 
Debates are fueled by the ever-improving quality of the observations. For 
instance, high spatial resolution and high temporal cadence 
observations reveal that oscillations in intensity and Doppler velocities 
of chromospheric lines are suppressed in the vicinity, but not, or less, directly 
above quiet-Sun network elements. \citet{judge01}, using SUMER UV spectral
line time-series and TRACE continuum observations, referred to this phenomenon
as \textit{magnetic shadow} ``since the network element appears to
`cast a shadow' over the neighboring ... internetwork region". These authors 
offered several explanations for the formation of the magnetic shadow, favoring
mode conversion of MHD waves if the plasma-$\beta$ (ratio of
thermal to magnetic pressure) was about unity at the formation height of the 
shadow.

In a subsequent study, \citet{mcintosh01} used potential field extrapolations 
to reconstruct the magnetic field of the observed solar region, from which 
they concluded that the magnetic shadow is caused by
closed magnetic field arches and not by mode conversion of MHD waves, and
that the $\beta=1$ layer is likely located at the heights from which the 
UV continuum emerges. \citet{krijger01} also argued that at the observed height, 
mode conversion is unlikely to cause the magnetic shadow because this process
would only become important at greater heights.

Observations by \citet{vecchio07} of the chromospheric spectral line of 
\ion{Ca}{ii} $854.2$~nm, 
acquired with the Interferometric Bidimensional Spectrometer (IBIS) 
\citep{cavallini06}, most clearly revealed the ring and shadow-like appearance 
of the suppressed oscillatory power around quiet-Sun magnetic network patches
and at frequencies above the acoustic cut-off frequency. More recently,
\citet{kontogiannis+al2010a} showed similar results from H$\alpha$ profiles
obtained with the Dutch Open Telescope, revealing `the discrete role of the 
magnetic field ... which guides or suppresses the oscillations, depending on 
its inclination'. \citet{kontogiannis+al2010b} complemented this study with
a potential magnetic field extrapolation and concluded that the magnetic
shadow is associated with the $\beta < 1$ environment.

To shed light on the formation mechanism of the magnetic shadow, we applied 
detailed three-dimensional MHD simulations of magneto-acoustic waves 
to a model that shows both magnetic network and internetwork regions, 
although on a considerably smaller scale than the observations. 
We followed the simulation approach of \citet{nutto+al2012}, which allows for the 
visualiziation of magneto-acoustic waves as they propagate across a complexly 
structured, dynamic solar atmospheric model.
For the terminology used in this paper, we refer the 
reader to the introductory section of \citet{nutto+al2012}.

\section{The background model}\label{Sect:model} 

The three-dimensional simulations 
were carried out with the CO$^5$BOLD-code
\citep{freytag+al2012}, which solves
the magnetohydrodynamic equations for a fully compressible gas
including radiative transfer and a realistic equation of state.

\subsection{Initial magneto-atmosphere}\label{sec:init-magn-atmosph}

The employed three-dimensional model extends from the surface  
layers of the convection zone throughout the photosphere to the middle 
layers of the magnetically
structured chromosphere, corresponding to a height range from
$z=-1420$~km to $z=1360$~km, where $z=0$~km is defined by the height
of mean optical depth unity, $\langle \tau\rangle =1$. Rosseland mean
opacities are used for the radiation transfer. The
horizontal outline of the box is quadratic with the dimensions
$4800$~km by $4800$~km.

The numerical grid consists of $120$ cells in each horizontal
direction. In the vertical direction, the grid-cell size gradually
decreases from $50$~km near the bottom to $20$~km in the atmospheric
layers. Periodic boundary conditions apply to each horizontal
direction, while the bottom and upper boundary are transmittent,
allowing acoustic waves to leave the box without major reflection. 
The boundary conditions force the magnetic field to be vertical 
at the top and bottom boundaries but allow it to move freely in the
horizontal direction.

\begin{figure}
\centering
\includegraphics[width=0.48\textwidth]{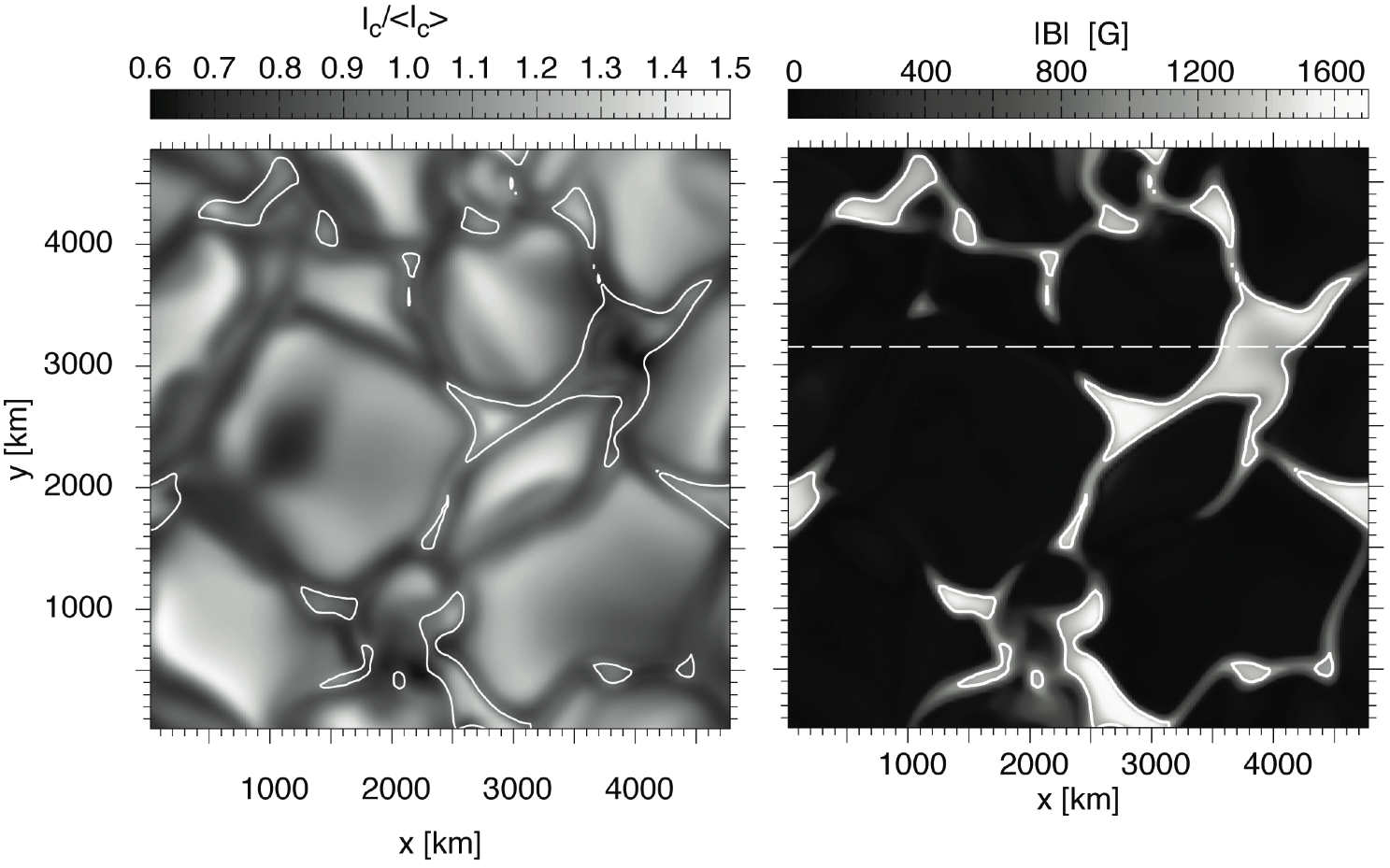}
%
\caption{Initial state. \textit{Left:} emerging radiative
  intensity at the top boundary. \textit{Right:} magnetic field
  strength on the $\tau=1$ surface of the initial
  model. The horizontal line indicates the position of the vertical
  slice shown in Fig.~\ref{fig:3d_dvz_10mHz}. The contours in both
  panels display the equipartition level where $c_s=c_A$.}
\label{fig:3d_initial_model}
\end{figure}

The left panel of Fig.~\ref{fig:3d_initial_model} shows the bolometric
intensity of the radiation that leaves the box through the top
boundary in the vertical direction. It shows the regular granulation
pattern of the solar surface and the intergranular lanes. The right
panel displays the absolute magnetic field strength of the initial
model at the surface layer of $\tau=1$.

This initial background model was derived from a simulation that
started with a vertically oriented, homogenous magnetic field of a
strength of $|\vec B|=200$~G. Because of the advective motion of the
granular cells, the magnetic field is concentrated in the
intergranular lanes.  In the course of the simulation, a strong flux 
concentration accumulates in the upper right quadrant of the box,
at $(x,y)=(3800,3100)$~km. Here, field strengths of up to 
$|\vec B_{\mathrm{max}}|=1600$~G are reached inside the flux tube at the
surface of $\tau=1$. The left half of the box contains a rather weak,
horizontally oriented magnetic field that resembles an internetwork region.

\subsection{Properties of the time-averaged model}\label{sec:prop-time-aver}

\begin{figure}
\centering
\includegraphics[width=0.48\textwidth]{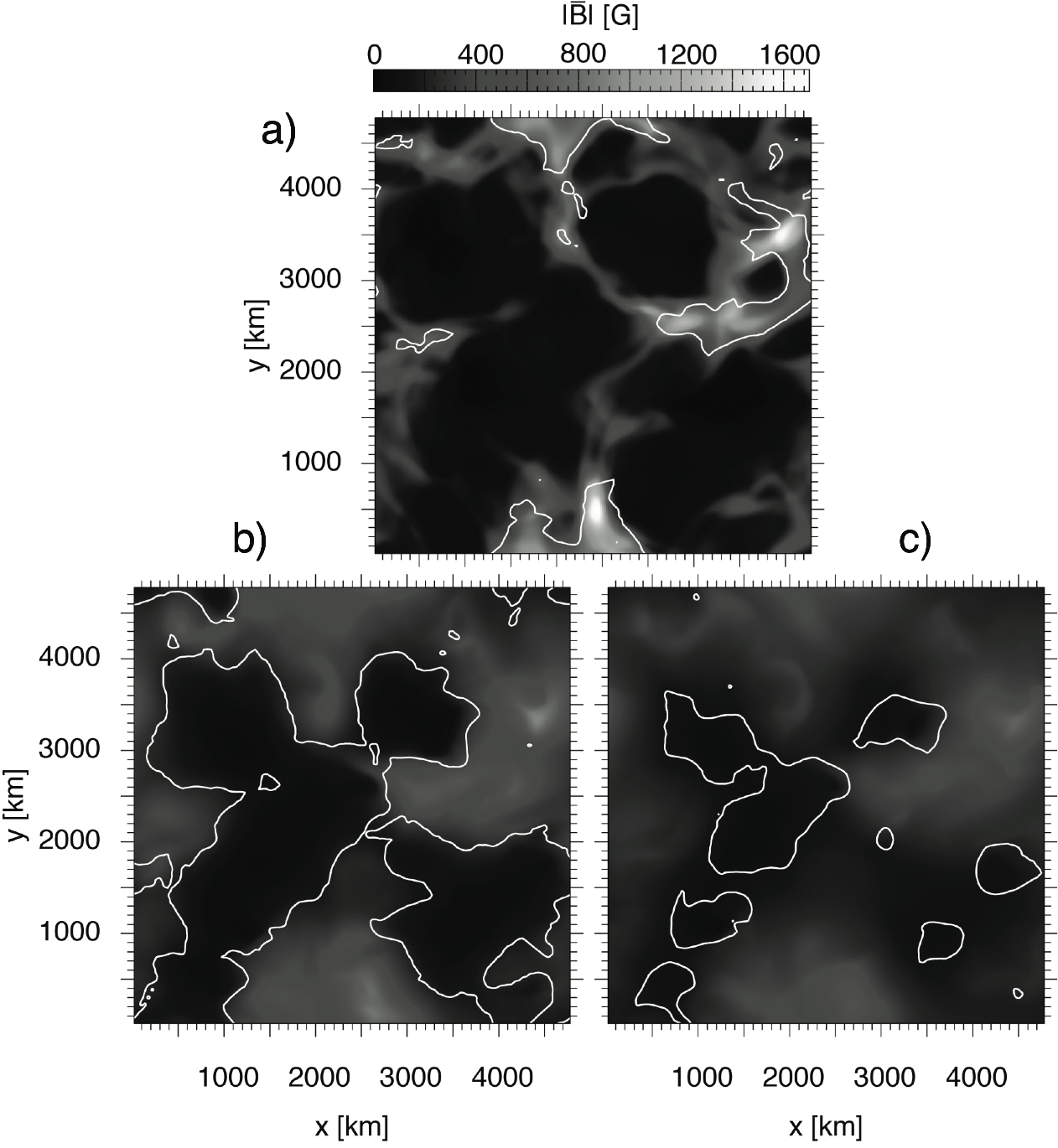}
\caption{Time-averaged absolute magnetic field strength, 
  $\overline{|\vec B|}$, at the 
  optical depth levels (a) $\tau_1=1$, (b) $\tau_2=0.0008$, and 
  (c) $\tau_3=6.7\times 10^{-5}$. The average
  is taken over the entire time series of $1250$~s. The horizontally
  averaged position of the considered optical depth levels corresponds
  to the geometrical heights, $\langle z(\tau_1)\rangle=0$~km,
  $\langle z(\tau_2)\rangle\approx 400$~km, and $\langle
  z(\tau_3)\rangle \approx 600$~km. The white contour represents the
  time-averaged position of the equipartition level where $c_A=c_s$.}
\label{fig:Bavg}
\end{figure}

Starting from the initial model, the simulations were advanced for
1250~s. Figure~\ref{fig:Bavg} shows the time-averaged (1250~s) 
magnetic field for different optical depth levels, which is
needed for the interpretation of the power maps that derive from the full length
of the time series.

The top panel of Fig.~\ref{fig:Bavg} displays the time-averaged absolute
magnetic field strength, $\overline{|\vec B|}$, on the surface of the 
time-averaged position of optical depth $\tau=1$. Compared to the map of 
the initial model shown in Fig.~\ref{fig:3d_initial_model}, the time-averaged
magnetic field, especially the strong flux concentration, appears
diffuse because of its continuous advective motion. Nonetheless, 
the time average still shows several
strong magnetic field concentrations of about $1600$~G. At the
transition to chromospheric levels, at $\tau_2$ and $\tau_3$, the
magnetic canopies of several magnetic field concentrations merge and
form a magnetically dominated chromosphere
\citep[cf.][]{solanki90}. Below the magnetic canopy, the time-averaged
background model contains a fairly weak, horizontally oriented magnetic
field, similar to the initial model.

\section{Wave propagation}\label{sec:wave-propagation}

Below, we show snapshots of the wave propagation
simulations. We use the subtraction method described by \citet{steiner07}
for the visualization in which physical quantities of two
simulation runs with and without an artificially introduced perturbation
are subtracted from each other. Test results for this approach
are given in \citet{nutto+al2010}.

The wave excitation is analogous to the two-dimensional case presented
in \citet{nutto+al2012}, where plane-parallel waves are excited at the
lower boundary. The same was performed for the three-dimensional case,
except that the wave was now excited along the entire plane of the
lower boundary. The applied driver excites monochromatic
plane-parallel waves with a frequency of $10$~mHz. The waves are
continuously excited for the whole duration of the simulation of
$1250$~s. The simulation can only be advanced for a limited
amount of time after which the waves become swamped in the background
of the residual large-scale velocities \citep[see][]{nutto+al2012}.
Therefore, we choose the relatively high frequency of $f_0 = 10$~mHz to
increase the reliability of the analysis for the available limited time 
period and also for convenient visualization of the waves.
One can expect a similar behavior of propagating waves for 
all frequencies above the acoustic cut-off frequency
\citep{khomenko06}, therefore waves with $f_0=10$~mHz should be a good 
representation for all waves above the acoustic cut-off frequency of 
$\approx 5$~mHz.

\begin{figure}
\centering
\includegraphics[width =.38\textwidth]{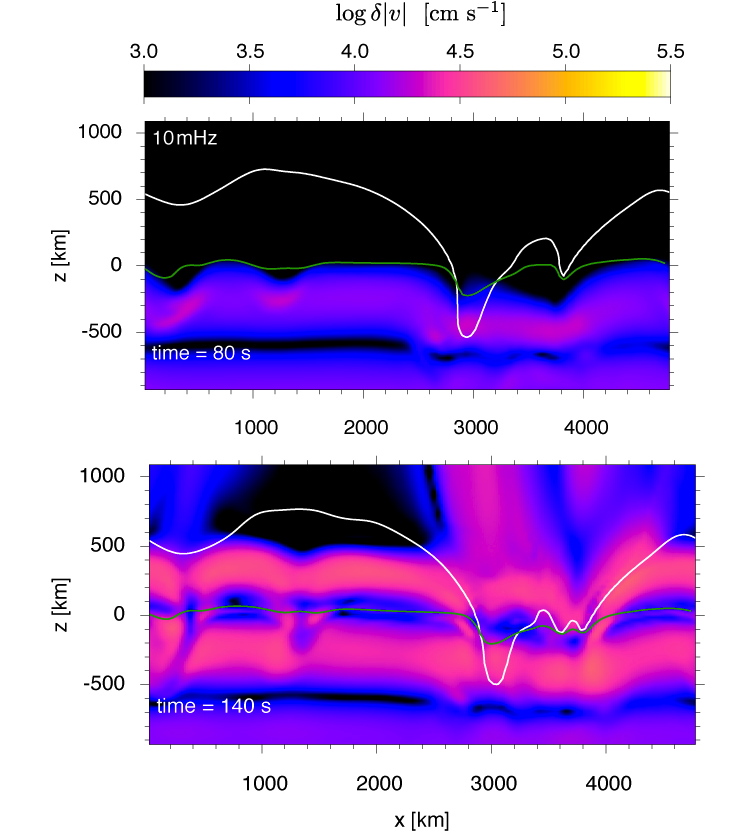}
\caption{Time instances of the wave-propagation simulations showing
  the logarithm of the absolute velocity perturbation, 
  $\delta |\vec v|$. Time increases from top to bottom. The vertical 
  slice is taken along the $xz$-plane at the position of a strong 
  magnetic flux tube (see horizontal line in the right panel of
  Fig.~\ref{fig:3d_initial_model}). The green contour corresponds to
  optical depth $\tau=1$. The white contour displays the equipartition
  level.}
\label{fig:3d_dvz_10mHz}
\end{figure}

At the bottom of the simulation domain and throughout the simulated
portion of the convection zone, we have $\beta \gg 1$ so that the excited 
wave is a fast predominantly acoustic mode. The upper panel of
Fig.~\ref{fig:3d_dvz_10mHz} shows the wave as it reaches the
solar surface. Mode conversion has not significantly taken place yet, 
because in the center of the flux tube, the angle between the wave vector 
and the magnetic field (attack angle) favors the transmission of the 
acoustic mode \citep{cally07}. 

In the lower panel of Fig.~\ref{fig:3d_dvz_10mHz}, the wave has 
traveled through the photosphere and
has reached the lower layers of the chromosphere. Clearly, the wave
field shows a different propagation behavior inside and outside the
magnetic canopy of the flux concentration.
Below the magnetic canopy, and hence between the magnetic network
elements, the wave propagation is nearly undisturbed. For regions
inside the magnetic network elements, viz., above the equipartition
level\footnote{The equipartition level is defined as the location
  where the Alfv\'en and the sound speed are of equal magnitude,
  $c_A=c_s$}, the wave fronts are vertically directed and belong 
to the fast magnetic mode that is excited because of mode conversion at 
the equipartition level. Owing to steep gradients of the Alfv\'en speed 
in the vertical and horizontal directions, the fast magnetic mode is
strongly refracted. In the apex of the refractive path, the fast
magnetic mode travels horizontally with the wave front being aligned
with the vertical direction.

In the very center of the magnetic flux element and coexistent with the
vertical wave fronts of the fast magnetic mode, we see the transmitted
slow acoustic mode. Its wave front keeps the horizontal
orientation from below and travels vertically upward along the
magnetic field lines of the flux concentration, reaching the higher
layers of the solar atmosphere much later than the fast mode.

\section{Power maps}\label{sec:power-maps}

We restricted our investigation to power maps of the vertical velocity
perturbations, $\delta v_z$, because these can be directly compared to
power maps of line-of-sight velocities observed at the center of the
solar disk.

The power maps were taken at constant optical depth levels $\tau_1=1$,
$\tau_2=0.0008$, and $\tau_3=6.7\times 10^{-5}$ in the atmosphere. The
time-averaged positions of these optical depth levels correspond to the
geometrical heights $\langle z(\tau_1)\rangle=0$~km,
$\langle z(\tau_2)\rangle\approx 400$~km, and
$\langle z(\tau_3)\rangle\approx 600$~km. The time series was
sampled with a cadence of $\Delta t=1$~s.
The power of observed Doppler shifts of active regions are usually
normalized to regions of the quiet Sun that show the least magnetic
activity. In our simulations, where the whole simulation domain is
threaded by a magnetic field, we cannot determine a corresponding
reference value. Therefore, we normalized the power to the value of the
spatially averaged power of each optical depth level.

\begin{figure*}
\centering
\includegraphics[width =.77\textwidth]{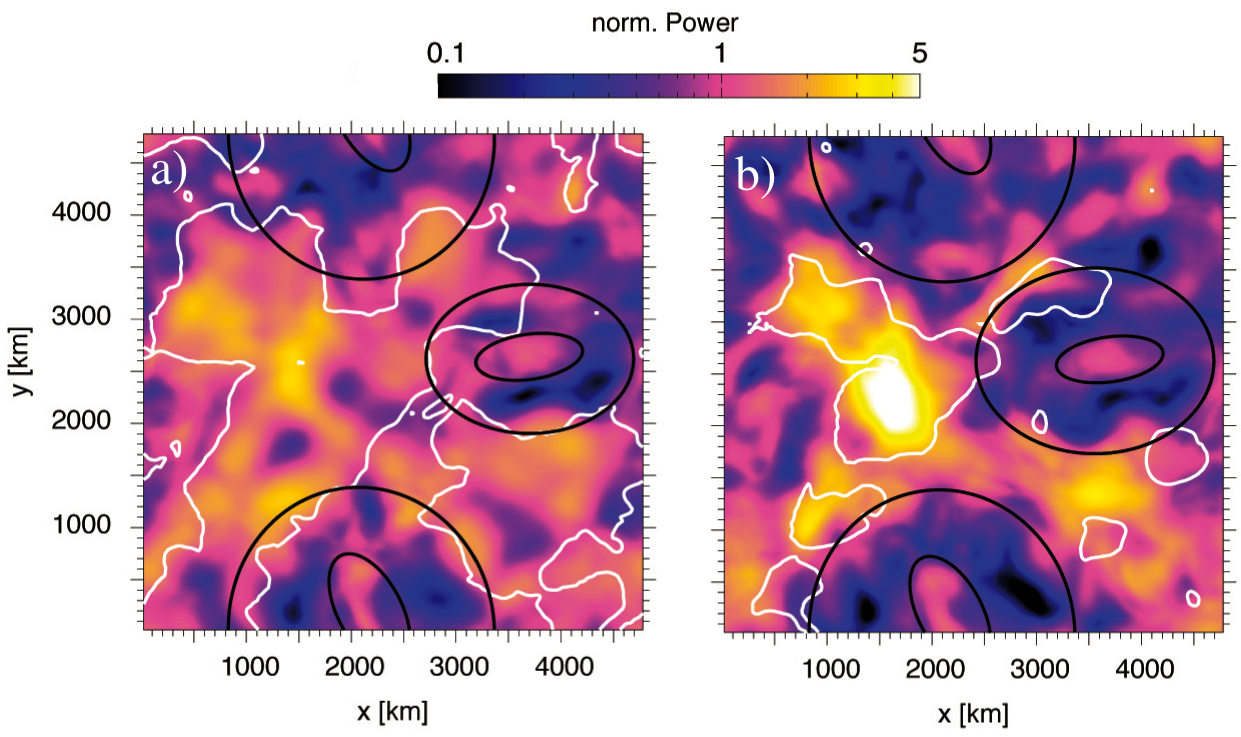}
\caption{Power maps of the vertical velocity perturbations, 
  $\delta v_z$, taken at the photospheric optical depth levels 
  $\tau_3=0.0008$ (a), and at the lower chromospheric layer, 
  $\tau_4=6.7\times 10^{-5}$ (b).  The white contours show the 
  time-averaged position of the equipartition level. The ellipses 
  mark regions where the \textit{magnetic shadow} can be identified. 
  Between the large and the small ellipses, the characteristic ring of 
  suppressed power is apparent (cf. Fig.~2 of \citet{vecchio07}).}
\label{fig:powermaps}
\end{figure*}

The derived power maps for the upper photosphere, $\tau_2$, and the
lower chromosphere, $\tau_3$, presented in Fig.~\ref{fig:powermaps},
yield rings of suppressed power (indicated by ellipses) around strong
magnetic field elements. These resemble very much the magnetic shadows
observed by \citet{vecchio07} in the lower chromosphere above magnetic
network elements. Three separate regions of different oscillatory
strength can be identified. The strongest oscillatory power occurs in
the internetwork regions (areas inside the equipartition contour
where $\beta >> 1$), while intermediate power can be found in the
very center of the magnetic network elements (inner ellipses). The
suppressed oscillatory power, and hence the shadow-like appearance,
occurs in the regions between the inner and outer ellipses.

The variable strength of the oscillatory power for the different
regions is caused by the presence of various modes of the
magneto-acoustic wave. In the internetwork regions the strong power
results from the nearly undisturbed fast acoustic mode which propagates
upward through the almost field-free medium
(cf.~Fig.~\ref{fig:3d_dvz_10mHz}). Here, the wave mode primarily
possesses a longitudinal polarization with velocity perturbations,
$\delta v_z$, along the propagation path and line-of-sight. 
Thus, the full strength of the oscillation shows up in the power map.

In the very center of the magnetic flux tubes, where the wave vector
and the magnetic field vector are parallel, the fast acoustic mode has
been primarily transmitted to the slow acoustic mode. This upwardly
propagating slow acoustic mode of longitudinal polarization causes the
intermediate oscillatory power in the very center of the magnetic
network elements. Compared to the power of the acoustic mode in the
internetwork region, the reduction of power is owed to the mode
conversion process, where a fraction of the acoustic energy has been
converted into the fast magnetic mode.

As shown in Sect.~\ref{sec:wave-propagation}, the fast magnetic mode 
propagates in the region between the internetwork and the center of 
the network elements. The suppression of the oscillatory power can now be
understood in terms of wave polarization---specifically by the orientation 
of the wave vector with respect to the magnetic field and line-of-sight. 
In the apex of the refractive wave path, the fast magnetic mode travels basically perpendicular to the magnetic field with longitudinal wave polarization and
consequently, with velocity perturbations in the direction of the horizontally 
directed wave vector. Therefore, along the propagation path, the initial vertical 
velocity perturbations, $\delta v_z$, are gradually transformed into horizontal 
velocity perturbations, $\delta v_{x,y}$, and back again. Thus, the 
oscillatory power of the vertical velocity
perturbations, $\delta v_z$, corresponding to line-of-sight velocities
at the solar disk center, are suppressed around magnetic network elements,
causing the appearance of a magnetic shadow.

According to \citet[and references therein]{khomenko12}, 
part of the fast 
magnetic mode can convert into an Alfv\'en wave, which then propagates along 
the magnetic field lines. The fast-to-Alfv\'en conversion takes place most 
efficiently in the evanescent tail of the fast wave beyond
the apex of the refractive wave path (reflection point), which extends 
throughout the upper chromosphere and beyond \citep{khomenko12}. 
Therefore it cannot be expected to leave imprints on 
the power maps of Fig.~\ref{fig:powermaps}, which pertain to the low 
chromosphere and upper photosphere, although it may become detectable in
the core of H$\alpha$. Moreover, in our case, the apex of 
the refractive wave path occurs in regions threaded by mainly vertically
directed magnetic fields. Hence, the excited Alfv\'en waves would not 
significantly contribute to line-of-sight velocity perturbations because of 
their transversal character. Consequently, we can consider the generation and 
propagation of the fast magnetic mode as the primary cause for the appearance 
of the magnetic shadow.

\section{Conclusion}\label{sec:conclusion}

Power maps of high-frequency waves observed in chromospheric spectral lines
show a reduction in power in the surroundings of network magnetic patches
reminiscent of a shadow. The responsible physical process for this appearance 
has remained 
ambiguous so far. With three-dimensional radiation magnetohydrodynamic simulation, 
we can now reproduce the magnetic shadow phenomenon and we conclude that it is
tied to the process of mode conversion. The excitation of the fast magnetic
mode in the expanding region of the network magnetic field and its refractive wave
path reduces the oscillatory power of the vertical velocity perturbations, 
$\delta v_z$, in the surroundings of magnetic network elements.

Furthermore, we can show that the power maps taken in the upper
photosphere and lower chromosphere show the oscillatory signatures of
different magneto-acoustic wave modes in three different
phenomenological regions: the fast acoustic mode in the internetwork,
the fast magnetic mode with possible superposition of Alfv\'en waves
in the magnetic canopy region, and the slow acoustic mode in the center 
of magnetic network elements.

\begin{acknowledgements}
  The authors acknowledge support from the European Helio- and
  Astroseismology Network (HELAS), which is funded as Coordination
  Action by the European Commission's Sixth Framework Programme.
  We thank the referee for the thorough reading of the manuscript
  and for constructive remarks.
\end{acknowledgements}

\bibliography{nutto+al}

\end{document}